\documentclass[12pt,a4paper]{article}
\usepackage{epsfig,amsmath,amssymb,subfigure,placeins}
\usepackage[colorlinks=true,linktocpage=true,linkcolor=blue,citecolor=blue]{hyperref}
\usepackage{color}
\newcommand{\be}{\begin{eqnarray}}
\newcommand{\ee}{\end{eqnarray}}
\newcommand{\qhat}{\hat{q}}
\begin{document}
\begin{center}
{\large{\bf Quenching parameter in a holographic thermal QCD}\\

Binoy Krishna Patra \footnote{binoyfph@iitr.ac.in} and Bhaskar Arya $^2$,
\\

{\it $^1$Department of Physics, Indian Institute of 
Technology Roorkee, Roorkee 247 667, India} \\
{\it $^2$ Department of Mechanical and Industrial Engineering, 
Indian Institute of Technology Roorkee, Roorkee 247 667, India} }
\end{center}

\begin{center}
Abstract
\end{center}
We have calculated the quenching parameter, $\hat{q}$ in a model-independent 
way using the gauge-gravity duality. In earlier 
calculations, the geometry in the gravity side at finite temperature was 
usually taken as the pure AdS blackhole metric for which the dual gauge theory 
becomes conformally invariant unlike QCD. Therefore we use
a metric which incorporates the fundamental quarks by embedding the 
coincident D7 branes in the Klebanov-Tseytlin background
and a finite temperature is switched on by inserting  a black 
hole into the background, known as OKS-BH metric. Further 
inclusion of an additional UV cap to the metric prepares the 
dual gauge theory to run similar to  thermal QCD.
Moreover $\hat{q}$ is usually defined in the literature 
from the Glauber-model perturbative QCD evaluation of the Wilson loop,
which has no reasons to hold if the coupling is large and
is thus against the main 
idea of gauge-gravity duality. Thus we use an appropriate definition of
$\qhat$: $\hat{q} L^- = 1/L^2$, where $L$ is the  
separation for which the Wilson loop is equal to some specific value.
The above two refinements cause $\qhat$ to vary with the temperature
as $T^4$ always and to depend linearly on the light-cone time $L^-$ 
with an additional ($1/L^-$) correction term in the short-distance limit
whereas in the long-distance limit, $\qhat$ depends only linearly on 
$L^-$ with no correction term. These observations agree with other 
holographic calculations directly or indirectly.
\section{Introduction}
In the initial stage of ultrarelativistic heavy-ion collisions energetic 
partons in the form of jets are produced from the hard collisions. After 
receiving a large 
transverse momentum, these jets plough through the fireball for a 
transitional period of about a few ${\rm{fm}}/c$ 
and will thus loose energy due to the interaction of the 
hard partons with the medium constituents, known as the jet 
quenching. As a result
the yield of hadrons with high transverse momentum ($p_T$) 
is shown to be significantly suppressed in comparison with the 
cumulative yields of nucleon-nucleon collisions. 
There are mainly two contributions to the energy loss of the partons in the
medium: one is due to the radiation emitted by the decelerated colour 
charges, {\em i.e.} bremsstrahlung of gluons~\cite{Baier:NPB483'1997, 
Zakharov:JETPLETT65'1997, Gyulassy:NPB594'2001}
and the other one is due to the collisions among the partons in the 
medium~\cite{Mustafa:APHA22'2005}.

The experimental discoveries at RHIC revealed that the matter 
produced is a strongly coupled quark-gluon plasma (sQGP) unlike
weakly interacting gas of partons expected from the naive asymptotic
freedom, {\em for example}, the observed elliptic flow,
the quenching of jets while traversing through the medium etc. The 
jet quenching is parametrized by the quenching parameter, $\qhat$, which 
is defined by the average transverse momentum square transferred from the 
traversing parton per unit mean free path. The
extracted values of this transport coefficient in realtivistic heavy-ion 
collisions by the JET collaboration~\cite{Bruke:PRC90'2014} range 
from 1-25 ${\rm{GeV}^2}$/fm, which are much
larger than those estimated from the perturbative QCD calculations.
This hints some non-perturbative mechanisms which may contribute to the
jet quenching mechanism. Thus it is
worthwhile to calculate the possible values of $\qhat$ 
in the strong coupling limit. The first 
principle lattice QCD however, cannot be applied for this 
purpose, which requires the real-time dynamics. 

The simplest gauge-gravity duality~\cite{Maldacena:PRL80'1998,Gubser:PLB428'1998,Witten:ATMP2'1998} between
the type IIB superstring theory formulated on AdS$_5\times S^5$ space 
and $\mathcal N$=4 supersymmetric
Yang-Mills theory (SYM) in four dimensions
provides a robust tool to explore 
the thermodynamical and transport properties of sQGP.
Although the underlying dynamics, QCD is 
different from $\mathcal N=4$ SYM but the correspondence seems feasible
because some of the properties of all strongly interacting systems 
show some universality behavior. One of the notable observation is the universal value,
($1/{4\pi}$) for the $\eta/s$ ratio for the quantum field
theories having a holographic description~\cite{Son:ARNPS57'2007} and thus 
it gives a lower bound to the ratio for sQGP.
Motivated by these similarities between the $\mathcal N=4$ SYM and 
the corresponding theory of supergravity, 
the jet-quenching parameter, $\qhat$ was related to the expectation 
value of the Wilson loop $W^A[{\cal C}]$ in adjoint 
representation due to the Eikonal approximation~\cite{Liu:PRL97'2006}:
\begin{equation}
\langle W^A[{\cal C}] \rangle  \approx \exp \left( -\frac{1}{4\sqrt{2}}
\hat{q}L_-L^2 \right)~,
\label{qpara}
\end{equation}
where ${\cal C}$ is a rectangular contour of size $L\times L_-$, with the 
sides, having the length $L_-$ run along the light-cone. 
There were other calculations of $\hat{q}$
~\cite{AHMueller:NPB335'1990,Dominguez:NPA811'2008,Casalderrey:NPA'2014}
using a very different setup and arriving at different conclusions. 
In the context of relativistic heavy ion collisions,
the effects of finite t'Hooft coupling ($\lambda$) as well as
chemical potential on $\hat{q}$ was studied in
~\cite{Armesto:JHEP0609'2006,Lin:PLB641'2006,Li:arXiv:1605.00188}
and the jet stopping
in strongly-coupled QCD-like plasmas with gravity duals have also 
been studied using the string $\alpha^\prime$ expansion in AdS/CFT
\cite{Arnold:JHEP1302'2013,Arnold:JHEP1207'2012}.

However, since $\qhat$ is related to the transverse momentum 
($p_T$) broadening  so to calculate the mean $p_T$, we need 
to Fourier transform (FT) of the Wilson loop
\be
W(p_T) = \int d^2 L~e^{i p \cdot L}~W(L) .
\ee
The above FT emerges if we intend to calculate the particle production in 
the scattering of a quark on a target, and the target will be the medium 
in the jet quenching problem. It turns out that the above FT 
is proportional to the quark production cross section, 
$W(p_T) \sim d\sigma/d^2 p$~\cite{Eramo:PRD84'2011,Abir:PLB748'2015}. 
Let us explore the subtleties which might help
us to search for the correct definition of $\qhat$. {\em For example}, 
if we define $\qhat$ as $\langle p_T^2 \rangle/L^-$, as some authors do. So
we would then need to find $\langle p_T^2 \rangle$. But this seems easy 
because $ \langle p_T^2 \rangle \sim \nabla_\perp^2 W(L)$ at L=0. This 
seems consistent 
with getting the coefficient of the $L^2$ term in the exponent, as in 
(\ref{qpara}). However, since our aim is to model QCD and in QCD at high 
$p_T$ perturbative physics works, and $d\sigma/d^2 p \sim 1/p_T^4$, so
$\langle p_T \rangle$ is infinite (irrespective of what happens at lower 
$p_T$). In other words one cannot trust $W(L)$ from AdS at very small $L$. 
A way out is to define $\qhat$ as ${\langle p_T \rangle}^2/L^-$. Since 
$\langle p_T \rangle $ is finite even in perturbative QCD, this definition 
is safe. To find $ \langle p_T \rangle $ we need the typical momentum scale 
of $W(L)$ and if one knows $W(p_T)$, then one should be able to find 
$\langle p_T \rangle $ exactly. Otherwise one could argue that 
$\langle p_T \rangle$ is given by the saturation scale $Q_s$, as the 
only scale available in the problem at high enough energy. Hence the standard 
prescription of finding $Q_s$ by requiring the Wilson loop, $W (L=1/Q_s)$ 
to be a constant, should 
probably give one a good estimate of $ \langle p_T \rangle $.

In summary the above definition of  $\hat{q}$ in (\ref{qpara}) as a 
coefficient of the $L^2$ term in the 
Wilson line correlator may not be correct because the motivation for the
definition (\ref{qpara}) in~\cite{Liu:PRL97'2006} comes from 
the Glauber-model 
perturbative QCD evaluation of the Wilson loop~\cite{Albacete:JHEP0807'2008,
Kovchegov:Cambridge'2012}.\footnote {In fact, it is already incorrect once 
someone includes perturbative QCD corrections to the Glauber formula.}
Therefore this perturbative expression has no reasons to hold when the
coupling is large, which is the main idea of gauge-gravity duality. A more 
appropriate
definition of $\hat{q}$ is then to postulate the equation
\be
\hat{q} L^- = 1/L^2~,
\label{qhatourdef}
\ee
where $L$ is the quark-antiquark separation for which the expectation value 
of the Wilson loop in adjoint representation is equal to some specific 
value. The above definition (\ref{qhatourdef}) can also be understood as 
follows: since $\hat{q} L^-$ behaves like the saturation scale squared in 
small-$x$ physics and the saturation scale is defined by requiring that 
the expectation value of Wilson loop is equal to some constant at 
$L = 1/Q_s$~\cite{Kovchegov:Cambridge'2012,Albacete:JHEP0807'2008}.

The calculations for $\qhat$ discussed so far used the geometry as the 
pure AdS black hole metric, for  which the dual gauge theory 
is conformally invariant SYM theory {\em unlike the QCD}.
This is one of the central theme of our work. Therefore, the aim of the 
present paper is to extend/modify the shortcomings of
the abovementioned calculation~\cite{Liu:PRL97'2006} in two fold: (i) the first
aim is to study the jet quenching in a gravitational background which is 
dual to a gauge theory 
with a RG flow that confines in the far IR and is asymptotically free at the 
far UV.  Recently a gravity dual with a black hole and 
seven branes embedded via Ouyang embedding is 
constructed~\cite{Mia:NPB839'2010,Mia:PRD82'2010}, which resembles
the main features of strongly coupled QCD, {\em i.e.} is almost conformal 
in the UV with no Landau poles or UV divergences of the Wilson loops, but has
logarithmic running of coupling in the IR.
Recently one of us have explored the properties of heavy quarkonium bound
states with the above geomtry and the 
findings~\cite{Binoy:PRD91'2015,Binoy:PRD92'2015} can only be 
understood as the artifact of the correct geometry for real QCD.
(ii) The second one is the appropriate definition
of $\qhat$ as in (\ref{qhatourdef}) for which the Wilson
loop is equal to some specific value, {\em say} 1/2. Our work is therefore
organized as follows. Section 2 will be devoted to revisit 
the Ouang-Klebanov-Strassler geometry and its improvements at 
the UV sector. In Section 3.1, we employ the aforesaid geometry to
obtain the renormalized Nambu-Goto action in both short- and long-distance
limits. Thereafter we will obtain the quenching parameter 
in Section 3.2 and will also discuss briefly the results of other 
calculations. Finally we conclude in Section 4.

\section{Construction of dual geometry}
A conformal gauge theory does not flow with the scale, hence
it has a trivial RG flow.
The AdS/CFT correspondence conjectures that a conformal theory in four 
dimension can be mapped on the boundary of a pure anti-de Sitter 
space~\cite{Maldacena:PRL80'1998}. But if the theory has 
a non-trivial RG flow like QCD, which is 
confining in IR and conformal in UV, we 
cannot describe the full theory on the boundary of some higher 
dimensional space and hence need to envisage differently at 
running energy scales. One way out is to embed the D branes in the 
geometry and as a result
the corresponding gauge theory exhibits logarithmic RG flow. 
Such a construction was done in the Klebanov-Strassler (KS) 
geometry~\cite{Klebanov:JHEP0008'2000} through a warped deformed conifold with three-form 
type IIB fluxes and the corresponding dual gauge theory is confining 
in the far IR limit but is not free at UV limit. The other demerits 
of the KS geometry are that it is devoid of quarks in the fundamental 
representation and cannot be generalized to finite temperature. 

The inclusion of fundamental matter in string theory is 
possible by embedding a set of flavor branes in  addition to the color 
branes. The strings connecting 
to the color and flavor branes in the adjoint representation of 
U($N_c$) group give the gauge particles and the mesons, respectively
whereas those connected to both the flavor 
and color  branes in the fundamental representation give the quarks 
and anti-quarks, 
respectively. In principle one could go to large number of color ($N_c$) 
and flavor $N_f$) branes in the near horizon limit and translates
the branes into fluxes and then construct the gravity background
which is holographically dual to gauge theory of quarks and gluons.
In practice the back reaction of the probes 
on the background could be neglected through the probe approximation
($ N_f \ll N_c$) and the flavor physics is then extracted 
by analyzing the effective action which describes the flavor 
branes in the color background\cite{Karch:JHEP0206'2002,Sakai:JHEP0309'2003}.
Since the full global solution for the backreaction of
D7 branes in the KS background becomes nontrivial so the insertion of the 
fundamental quarks in the original KS geometry~\cite{Klebanov:JHEP0008'2000} becomes
difficult. Peter Ouyang~\cite{Ouyang:NPB699'2004} has successfully 
put the coincident D7 branes into the 
Klebanov-Tseytlin background~\cite{Klebanov:NPB578'2000}, known as OKS
geometry, which has all the type IIB fluxes switched on including the 
axio-dilaton and the local metric was then computed by incorporating 
the deformations 
of the seven branes by moving them far away from the regime of interest.
Hence the axion-dilaton vanishes for the background locally, but there will be
non-zero axion-dilaton globally, as a result the local back reactions
on the metric modify the warp factors to the full global scenario.

For realizing the fnite temperature a black hole is inserted into 
the OKS background, {\em i.e.} OKS-BH geometry, where
the Hawking temperature corresponds to the gauge theory temperature.
Thus the metric in OKS-BH geometry is expressed in terms of warp factor ($h$)
~\cite{Mia:NPB839'2010} 
\begin{equation}
ds^2 = \frac{1}{\sqrt{h}}
\Big[-g_1(u)dt^2+dx^2+dy^2+dz^2\Big] +\sqrt{h}\Big[g_2^{-1}(u) du^2+ 
d{\cal M}_5^2\Big]
\label{m1}
\end{equation}
where $g_i(u)$ are the black-hole factors as a function of
the extra dimension, $u$ and $d{\cal M}_5^2$ is due to the warped 
resolved-deformed conifold. The gauge theory dual to the metric
(\ref{m1}) flows correctly at IR like QCD but the effective 
degrees of freedom grow indefinitely at UV limit. The situation becomes
worse even in the presence of fundamental flavors because its 
proliferation leads to Landau poles and hence the Wilson loops diverges at UV. 
To circumvent the problem, one need to add the appropriate 
UV cap to the AdS-Schwarzschild geometry in the asymptotic UV limit.
However, the additional UV caps, in general may deform the IR geometry but
the far IR geometry has not been changed because the UV caps correspond 
to adding the non-trivial irrelevant operators in the dual gauge theory.
These operators 
keep far IR physics completely unchanged, but the physics at not-so-small 
energies may be changed a bit. 

Recently the IR geometry part has been suitably modified to obtain the 
desired dual gauge theory by the McGill group~\cite{Mia:PRD82'2010,Mia:NPB839'2010,Mia:PLB694'2011}, where the metric (\ref{m1}) will receive further 
corrections, $g_{uu}$, because the unwarped metric may not remain Ricci flat
due to the presence of both axio-dilaton and seven-brane sources, as:
\begin{equation} \label{correction}
ds^2 = {1\over \sqrt{h}}
\Big[-g(u)dt^2+dx^2+dy^2+dz^2\Big] +\sqrt{h}\Big[g(u)^{-1}g_{uu}du^2+ g_{mn}dx^m dx^n\Big]
\end{equation}
where the black hole factors $g_i(u)$'s are set as $ g_1(u) = g_2(u) = g(u)$
and the corrections $ g_{uu} $ are of the form $1/u ^{n} $ and may be 
written as a series expansion:
\begin{equation}
g_{uu} = 1 + \sum_{i = 0}^{\infty} \frac{a_{uu,i}}{u^i}~,
\label{guu}
\end{equation}
where the coefficients, $ a_{uu,i}$ are  independent of the extra-dimension 
coordinate $u$ and are solved exactly in~\cite{Mia:PRD82'2010}. Thus the warp 
factor, $h$ can be extracted from the above corrections (\ref{guu}) as
\begin{equation*}
 h ~= ~\frac{L^4}{u^4}\left[1+\sum_{i=1}^\infty\frac{a_i}{u^i}\right]~~~ 
\label{warp}
\end{equation*}
where the coefficients, $a_i $ are of ${\cal O}(g_sN_f)$ and $L$
is the curvature of space. Thus the metric (\ref{correction}) reduces to 
OKS-BH in the IR limit and becomes $AdS_5 \times M_5$ in the UV limit, hence 
describes well both in IR and UV limits. Therefore, with the change 
of coordinates $ z = 1/u $, we can rewrite the metric (\ref{correction}) as
\be
ds^2 ~&=&~ g_{\mu\nu} dX^\mu dX^\nu \nonumber\\
&~=&~
 A_n z^{n-2}\left[-g(z)dt^2+d\overrightarrow{x}^2\right]
~ +~\frac{
B_l z^{l}}{
 A_m z^{m+2}g(z)}dz^2+\frac{1}
{A_n z^{n}}~ds^2_{{M}_5},
\label{OKSBH}
\ee
where $ds^2_{{\cal M}_5}$ is the metric of the internal space 
and the coefficients, $A_n$'s can be obtained from the 
coefficients, $a_i$ in the warp factor (\ref{warp})  as follows: 
\begin{equation}
{1\over \sqrt{h}}~ = ~ {1\over L^2 z^2 \sqrt{a_i z^i}} \equiv  
A_n z^{n-2} ~=~ {1\over L^2 z^2}\left[a_0  
-{a_1z\over 2} + \left({3a_1^2\over 8a_0} - {a_2\over 2}\right)z^2 
+ \cdot \cdot \cdot \right] \quad,
\end{equation}
which gives $ A _0 = {a_0\over L^2},  A_1 = -{a_1\over 2L^2},  
A_2 = {1\over L^2}\left({3a_1^2\over 8a_0} - {a_2\over 
2}\right)$ etc. Note that since $a_i$'s for  $i \ge 1$ 
are of ${\cal O}(g_sN_f)$ and $L^2 \propto \sqrt{g_sN}$, so 
in the limit $ g_s N_f 
\rightarrow 0$ and $ N \rightarrow \infty $ all $ 
A_i$'s for $ i \ge 1$ are very small. The second term in the 
metric (\ref{OKSBH}) accommodates the $1/u^{n}$ corrections 
in (\ref{correction}) via the series, $ B_l z^l$, which is 
expanded further:
\begin{equation}
B_l z^l = 1 + a_{zz,i} z^i.
\end{equation}

In an comprehensive study~\cite{Mia:NPB839'2010}, the entire geometry is 
split into three regions. Apart from the two asymptotic regions at IR and
UV, respectively, there is an interpolating region 3 where at the outermost 
boundary
the three-forms vanish and the innermost boundary will be the outermost 
boundary of region 1. The background in these three regions and the 
insertion of additional UV cap are extensively analyzed
by the corresponding RG flows and the
field theory realizations have been discussed in \cite{Chen:PRD87'2013}.
Recently another suitable model to study certain IR
dynamics of QCD is the Sakai-Sugimoto model~\cite{Sakai:PTP113'2005} in the 
type IIA string theory, which consists of a set of $N$ wrapped color 
D4-branes on the circle and the flavor branes D8 and $\bar{D}8$
placed at the anti-nodal points of the circle to conceive 
the mesonic bound states. In its dual gravity, the wrapped D4-branes
are replaced by an asymptotically
AdS space, but the eight-branes remain and so does the circular
direction. However the Sakai-Sugimoto model does not have a UV
completion and had been compared recently with the aforesaid 
gravity dual in~\cite{Dasgupta:JHEP1507'2015}.
We shall not go into the complete details here and will use the metric 
(\ref{OKSBH}) to obtain the Nambu-Goto action and hence the Wilson loop 
is computed through gauge-gravity correspondence in the next section. 

\section{Gauge-Gravity Duality}
According to the gauge/gravity prescription~\cite{Maldacena:PRL80'1998}, the
expectation value of the Wilson loop, $W(C)$ in a strongly 
coupled gauge theory is related to the generating functional of the
string in the bulk which has the loop $C$ at the boundary
\begin{equation}
\label{wilsongaugegravity}
\langle W(C) \rangle \sim Z_{\rm{string}} 
\end{equation}
In supergravity limit, the generating functional becomes 
\be
Z_{\rm{string}} =e^{iS_{\rm{string}}}~,
\ee
where $S_{\rm{string}}$ is obtained by extremizing the string action, known
as the Namu-Goto action. So the above correspondence (\ref{wilsongaugegravity}) 
is translated into 
\begin{equation}
\label{eq:wilsongaugegravity}
\langle W(C) \rangle \sim e^{i S_{\rm{string}}}
\end{equation}
Thus we will now evaluate the Nambu-Goto action in the next subsection.
\subsection{Nambu-Goto Action}
By the light-cone transformation, 
\be
dt&=&\frac{dx^++dx^-}{\sqrt{2}} \nonumber\\
d{x_1}&=&\frac{dx^+-dx^-}{\sqrt2}
\ee
the metric (\ref{OKSBH}) is rewritten in terms of light-cone coordinates
as 
\be
ds^2&=&\left[-\frac1{2}A_n{z^{n-2}}g+\frac1{2}A_n{z^{n-2}}\right]\left[{dx^+}^2
+{dx^-}^2\right]-\left(1+g\right)A_n{z^{n-2}}dx^+{dx^-} \nonumber\\
&+&A_nz^{n-2}\left[{dx_2}^2+{dx_3}^2\right]
+\frac{B_nz^n}{A_nz^{n+2}g}dz^2+\frac{1}{A_nz^n}~{ds_{M_5}}^2
\label{metriclight}
\ee
We parametrize the two-dimensional world sheet and their derivatives 
in terms of the light-cone coordinates 
\be
\tau &=& x^-,~~\sigma = x_2 \in [-\frac{r}{2}, \frac{r}{2}], \nonumber\\
x_2 &=& {\rm{const}}, ~ x_3 = {\rm{const}}, ~ z = z(x_2) \nonumber\\
\partial_{\alpha} &=& \frac{\partial}{\partial
\tau}, ~ \partial_{\beta} = \frac{\partial}{\partial \sigma}.
\label{param}
\ee
With above parametrization(\ref{param}), the elements of the induced 
metric defined by
\be
g_{\alpha\beta}=G{\mu\nu}\frac{\partial{x^\mu}}{\partial{\sigma^\alpha}}
\frac{\partial{x^\nu}}{\partial{\sigma^\beta}}
\ee
can be read off from the above metric (\ref{metriclight})
\be
g_{--}&=&\frac{{A_n{z^{n}}\left(1-g\right)}}{2z^2}\nonumber\\
g_{-2}&=&g_{2-}=0 \nonumber\\
g_{22}&=&\frac{A_n{z^{n}}}{z^2}+
\frac{B_nz^n}{z^2A_nz^ng}{z^\prime}^2~.
\ee
Thus the determinant of the induced metric, $g_{\alpha \beta}$  can be 
calculated
\be
\mathbf{det}~{g_{\alpha\beta}}=g_{--}g_{22}=\frac1{2{z_h}^4}
\left[\left(A_nz^n\right)^2
+\frac{\left(B_nz^n\right){z^\prime}^2}{g}\right]~,
\ee
hence the Nambu-Goto action can be obtained as
\be
S&=&-\frac{1}{2\pi{\alpha}^{\prime}}\int\int{d\sigma}d\tau 
\sqrt{-{\mathbf{det}} g_{\alpha \beta}}\nonumber\\
&=&-\frac{1}{2\pi{\alpha}^{\prime}}\int\int{d\sigma}d\tau 
\sqrt{-\frac1{2{z_h}^4}\left[\left(A_nz^n\right)^2+
\frac{\left(B_nz^n\right){z^\prime}^2}{g}\right]} ~,
\label{lagran}
\ee
where $\alpha^\prime$ (=$\frac{R^2}{\sqrt{\lambda}}$, $R$ is the AdS radius
and $\lambda$ is the t'Hooft coupling) is the string tension. Thus 
the equation of motion:
\be
z^\prime{\frac{\partial{{\cal L}}}{\partial{z^\prime}}}- {\cal L}=C 
\ee
can be written from the above Lagrangian (${\cal L}$) in (\ref{lagran})
as
\be
-\left(A_nz^n\right)^2=C\sqrt{\left(A_nz^n\right)^2+
\frac{\left(B_nz^n\right){z^\prime}^2}{g}}~,
\ee
where $C$ is a constant of motion and can be obtained from the 
condition: $z^\prime=0$ at $z=z_m$, 
\be
C^2=\left(A_nz_m^n\right)^2
\ee
After substituting the constant $C$, the equation of motion becomes
finally 
\be
{z^\prime}^2=\frac{\left(A_nz^n\right)^2g}{B_nz^n}
\left[\frac{\left(A_nz^n\right)^2}{\left(A_n{z_m}^n\right)^2}-1\right]
\label{eom}
\ee
Since the Lagrangian is independent of the time
so after integrating over the time-like coordinate ($x_-$), 
the action becomes
\be
S&=&-\frac{iL^-}{2\sqrt2\pi{\alpha}^\prime{z_h}^2}
\int_{-\frac{L}{2}}^{+\frac{L}{2}} dx_2
\sqrt{\left(A_nz^n\right)^2+\frac{\left(B_nz^n\right){z^\prime}^2}{g}} \\
&=&-\frac{i2L^-}{2\sqrt2\pi{\alpha}^\prime{z_h}^2}
\int_{0}^{z_m} dz
\sqrt{\frac{{\left(A_nz^n\right)}^2}{{z^\prime}^2}+
\frac{\left(B_nz^n\right)}{g}} 
\label{m5}
\ee
We will now substitute ${z^\prime}^2$ from the equation of 
motion (\ref{eom}) to obtain the action. Since 
$\left(A_nz^n\right)^2<<\left(A_n{z_m}^n\right)^2$ so neglecting
the higher-order terms and keeping up to the second-order term, 
the action (\ref{m5}) is simplified into 
\be
S \simeq -\frac{\sqrt2L^-}{2\pi\alpha^\prime{z_h}^2\left(1+A{z_m}^2\right)}
\int_0^{z_m} \frac{dz}{\sqrt{g}}~\left(1+\frac{B}2z^2\right)
\left(1+Az^2\right)
\label{action}
\ee
We will now evaluate the Nambu-Goto action by solving the above integral 
in both short- and long-distance limits :\\

\noindent Case-I: In the short-distance ($z_m<<z_h$) limit,
after performing the integration in (\ref{action}) the 
action is written in terms of Gaussian hypergeometric functions 
\be
S&=&-\frac{L^-}{\sqrt2\pi{\alpha}^\prime{z_h}^2\left(1+Az_m^2\right)}
\int_0^{z_m}dz\left(\frac{1+\frac{B+2Az^2}{2}+\frac{ABz^4}{2}}{\sqrt{1-\frac{z^4}{z_h^4}}}\right)\nonumber\\
&=&-\frac{L^-z_m}{\sqrt{2}\pi{\alpha}^\prime{z_h}^2(1+Az_m^2)}
\left[{}_2F_1 \left(\frac{1}{4},\frac{1}{2},\frac{5}{4};\frac{z_m^4}{z_h^4}\right)
+\frac{(B+2A)z_m^2}{6}~{}_2F_1 \left(\frac{1}{2},\frac{3}{4},\frac{7}{4};
\frac{z_m^4}{z_h^4} \right) \right.\nonumber\\
&\quad \quad \quad \quad \quad \quad +& \left. \frac{ABz_h^4}{6} \left(-\sqrt{1-\frac{z_m^4}{z_h^4}}+{}_2F_1\left(\frac{1}{4},
\frac{1}{2},\frac{5}{4};\frac{z_m^4}{z_h^4}\right)\right)\right]
\ee
On expanding the hypergeometric functions in powers of
($\frac{z_m}{z_h}$) 
\be
{}_2F_1 \left(\frac{1}{4},\frac{1}{2},\frac{5}{4};\frac{z_m^4}{z_h^4} \right)
&=&\left(1+\frac{{z_m}^4}{10{z_h}^4}+...\right) ~, \nonumber\\
{}_2F_1 \left(\frac{1}{2},\frac{3}{4},\frac{7}{4}; \frac{z_m^4}{z_h^4} \right)
&=&\left(1+\frac{3{z_m}^4}{14{z_h}^4}+...\right) ~,\nonumber\\
{}_2F_1 \left(\frac{1}{4}, \frac{1}{2},\frac{5}{4};\frac{z_m^4}{z_h^4} \right)
&=& \left(1+\frac{{z_m}^4}{10{z_h}^4}+...\right)~,
\ee
respectively and ignoring the higher-order terms beyond the second power,
the action becomes
\be
S \stackrel{z_{\rm{m}} \ll z_h}{ \simeq} -\frac{L^-z_m}{\sqrt{2}\pi\alpha^\prime z_h^2}\left[1+\frac{(B-4A)z_m^2}{6}+\frac{z_m^4}{10z_h^4}\right]
\label{actionshort}
\ee
In addition to the extremal surface constructed above for the Nambu-Goto
action, there is another trivial one given by the two disconnected world sheets, 
placed one at $x_2=+\frac{L}{2}$ and another at $x_2=-\frac{L}{2}$.
The action for these two surfaces is
\be
S_0&=&-\frac2{2\pi\alpha^\prime}\int{dzdx^-}\sqrt{-g_{--}g_{zz}}\nonumber\\
&=&-\frac{iL^-}{\sqrt{2}\pi\alpha^\prime z_h^2}\int_0^{z_m}{dz}~
\frac{1+\frac{B}{2}z^2}{\sqrt{1-\frac{z^4}{z_h^4}}} \label{actiontwostring}\\
&=&-\frac{iL^-z_m}{\sqrt{2}\pi\alpha^\prime z_h^2}\left[{}_2F_1\left(\frac{1}{4},
\frac{1}{2},\frac{5}{4};\frac{z_m^4}{z_h^4}\right)+\frac{Bz_m^2}{6}~
{}_2F_1\left(\frac{1}{2},\frac{3}{4},\frac{7}{4};\frac{z_m^4}{z_h^4}\right)
\right]
\ee
Expanding the above hypergeometric functions in powers of $(\frac{z_m}{z_h})$
\be
{}_2F_1 \left(\frac{1}{4}, \frac{1}{2},\frac{5}{4};\frac{z_m^4}{z_h^4}\right)
&=& \left(1+\frac{{z_m}^4}{10{z_h}^4}+...\right) ,\\
{}_2F_1 \left(\frac{1}{2},\frac{3}{4},\frac{7}{4};\frac{z_m^4}{z_h^4} \right)
&=& \left(1+\frac{3{z_m}^4}{14{z_h}^4}+...\right),
\ee
respectively and ignoring the higher-order 
terms beyond the second power, the action to be subtracted ($S_0$) becomes
\be
S_0 \stackrel{z_{\rm{m}} \ll z_h}{ \simeq} -\frac{iL^-z_m}{\sqrt{2}\pi\alpha^\prime z_h^2}\left[1+\frac{Bz_m^2}{6}+\frac{z_m^4}{10z_h^4} + \cdot \cdot \cdot \right]
\label{actionsub}
\ee
Therefore the renormalized action is obtained by subtracting the action
(\ref{actionsub}) for the two disconnected surfaces from(\ref{actionshort})
\be
S_I& \stackrel{z_{\rm{m}} \ll z_h}{ \simeq} &S-S_0\nonumber\\
&=&-\frac{L^-z_m}{\sqrt2\pi{\alpha}^\prime{z_h}^2}\left[\left(1+
\frac{\left(B-4A\right)z_m^2}{6}+\frac{z_m^4}{10z_h^4}\right)-i\left(1+\frac{Bz_m^2}{6}+\frac{z_m^4}{10z_h^4}\right)
\right]
\label{renactionshort}
\ee

\noindent Case II: In the long-distance limit ($z_m>>z_h$) the integral 
in the action (\ref{action}) is split into integrations:
\be
S&=&-\frac{L^-}{\sqrt{2}\pi \alpha^\prime {z_h}^2(1+A{z_m}^2)}
\left[\int_{0}^{z_h}dz
\frac{(1+\frac{Bz^2}{2})(1+Az^2)}{\sqrt{1-\frac{z^4}{{z_h}^4}}}+
\int_{z_h}^{z_m}dz
\frac{(1+\frac{Bz^2}{2})(1+Az^2)}{\sqrt{1-\frac{z^4}{{z_h}^4}}}\right] \nonumber\\
& \equiv & {\rm{I}} +{\rm{II}}~,
\ee
where the first integral (I) becomes
\be
{\rm{I}}&=&-\frac{L^-}{\sqrt{2}\pi \alpha^\prime {z_h}^2(1+A{z_m}^2)}
\left[\int_{0}^{z_h}dz
\frac{(1+\frac{Bz^2}{2})(1+Az^2)}{\sqrt{1-\frac{z^4}{{z_h}^4}}} \right]
\nonumber\\
&\simeq &-\frac{L^-}{\sqrt{2}\pi \alpha^\prime {z_h}^2}\left[1.3z_h
+0.3(B+2A){z_h}^3+0.22AB{z_h}^5\right]
\ee
and the second integral (II) becomes, after neglecting the higher-order terms
in powers of $(\frac{z_h}{z_m})$ and keeping up to the second order
\be
{\rm{II}}&=&-\frac{L^-}{\sqrt{2}\pi \alpha^\prime {z_h}^2(1+A{z_m}^2)} \left[
\int_{z_h}^{z_m}dz 
\frac{(1+\frac{Bz^2}{2})(1+Az^2)}{\sqrt{1-\frac{z^4}{{z_h}^4}}}\right] 
\nonumber\\
& \simeq &\frac{-iL^-}{\sqrt{2}\pi \alpha^\prime {z_h}^2}(1.14z_h+0.5(B+2A)z_m{z_h}^2+0.17AB{z_m}^3{z_h}^2)
\ee
Therefore the Nambu-Goto action in this limit becomes 
\be
S& \stackrel{z_{\rm{m}} \gg z_h}{ =} &-\frac{L^-}{\sqrt{2}\pi 
\alpha^\prime {z_h}^2}\left[(1.3z_h
+0.3(B+2A){z_h}^3+0.22AB{z_h}^5)\right.\nonumber\\
&+&\left. i(1.14z_h+0.5(B+2A)z_m{z_h}^2+0.17AB{z_m}^3{z_h}^2)\right]
\ee
Similarly the action to be subtracted (\ref{actiontwostring}) in this limit
can be written as 
\be
S_0 = 
-\frac{iL^-}{\sqrt{2}\pi\alpha^\prime z_h^2}\left[\int_0^{z_h}{dz}\frac{1+\frac{B}{2}z^2}{\sqrt{1-\frac{z^4}{z_h^4}}}+\int_{z_h}^{z_m} dz\frac{1+\frac{B}{2}z^2}{\sqrt{1-\frac{z^4}{z_h^4}}}\right]
\ee
After integrating and keeping the terms up to the second-order, the 
action, $S_0$ for two disconnected surfaces can becomes 
\be
S_0 \stackrel{z_{\rm{m}} \gg z_h}{ \simeq} -\frac{L^-}{\sqrt{2}\pi 
\alpha^\prime {z_h}^2}\left[-1.14z_h-0.3B{z_h}^3+i(1.3z_h+0.3B{z_h}^3)\right] \ee
Therefore, the renormalized action is given by
\be
S_I& \stackrel{z_{\rm{m}} \gg z_h} {=} &S-S_0\nonumber\\
&=&-\frac{L^-}{\sqrt{2}\pi \alpha^\prime {z_h}^2}\left[2.44z_h+0.5Bz_m{z_h}^2+i(-0.16z_h+0.5(B+2A)z_m{z_h}^2)\right]
\label{renoractionlong}
\ee
\subsection{Jet Quenching Parameter}
We will now obtain the quenching parameter, $\qhat$
for which the expectation value of the Wilson loop 
in the adjoint representation is equal to some specific 
value, say, $C$,
\be
 \langle W_A \rangle = e^{i2S_I}=C
\ee
In our problem, $\langle W\rangle$ becomes complex-valued, which is a feature previously encountered in
\cite{Albacete:JHEP0807'2008} as well. Since $\langle W \rangle$ is the 
S-matrix for a quark dipole-medium scattering, it is allowed to be 
complex. If we were calculating $Q_s$ we would need the imaginary 
part of the forward scattering amplitude: since $S = 1 + i T$, then
$\Im T = 1 - \Re S = 1 - \Re \langle W \rangle$. This was exactly done
in~\cite{Albacete:JHEP0807'2008}. Therefore we redefined $\qhat$ 
in (\ref{qhatourdef}), where $L$ is the
separation at which the real part of the Wilson loop is constant ($C$). 

Thus decomposing the renormalized action, $S_I$ into the real and
imaginary parts, the real part of the expectation value of Wilson loop is
\be
\Re \langle W_A \rangle &=&\Re\left[e^{i\left(2\Re S_I+2i\Im S_I\right)}
\right] \nonumber\\
&=&e^{-2 \Im S_I}\left[\cos\left(2\Re S_I\right)\right]=C
\label{wilsonconst}
\ee
Now we will evaluate the quenching parameter for both long- and 
short-distance limits, using the actions in the respective limits.\\

\noindent Case I: Short-distance limit ($z_m \ll z_h $)\\ 
To write the action as a function of the separation $L$, we first express
$z_m$ in terms of $L$. For that we rewrite the equation of motion
(\ref{eom}) in this limit ($z_m \ll z_h$)
\be
{z^\prime}^2=-\frac{\left(A_nz^n\right)^2g}{B_nz^n}
\label{eomshort}
\ee
because ${(A_nz^n)}^2$ is much less than ${(A_n{z_m}^n)}^2$. 
Integrating both sides of the equation of motion (\ref{eomshort})
\be
\int_0^{z_m}dz\frac{\sqrt{B_nz^n}}{(A_nz^n)\sqrt{g}}=i \int^0_{-L/2}dx_2
\ee
the separation (L) becomes
\be
\frac{iL}{2}&=&\int_0^{z_m}dz\frac{(1+0.5Bz^2)(1-Az^2)}{\sqrt{1-
\frac{z^4}{{z_h}^4}}} \nonumber\\
&=&z_m+\frac{z_m^5}{10z_h^4}+0.17(B-2A)z_m^3\left(1+\frac{3z_m^4}{14z_h^4}\right)\nonumber\\
&-&0.17ABz_h^4z_m\left(-\sqrt{1-\frac{z_m^4}{z_h^4}}+ 
{}_2F_1 \left( \frac{1}{4},\frac{1}{2},\frac{5}{4},\frac{z_m^4}{z_h^4}
\right) \right)
\ee
Inverting the series and ignoring the higher-order terms we can express 
$z_m$ as a function of $L$ as
\be
z_m=\frac{Li}{2}\left(1+\frac{\left(B-2A\right)}{24}L^2\right)
\ee
Thus the renormalized action can expressed in terms of the 
separation ($L$) by replacing $z_m$ as a function of
$L$ into (\ref{renactionshort}). Ignoring the higher-order terms, the 
renormalized action is then given by
\be
S_I&=&-\frac{\sqrt2L^-}{2\pi{\alpha}^\prime{z_h}^2}\frac{Li}{2}
\left(1+\frac{\left(B-2A\right)L^2}{24}\right)\left[\left(1+
\frac{\left(B-4A\right)L^2}{24}+\frac{L^4}{160z_h^4}\right) \right.\nonumber\\
&-i&\left. \left(1+\frac{BL^2}{24}+\frac{L^4}{160z_h^4}\right)\right]
\ee
Now the imaginary and real parts of the renormalized action can be
separated, respectively as
\be
\Im S_I=-\frac{\sqrt2L^-}{2\pi{\alpha}^\prime{z_h}^2}\frac{L}{2}
\left(1+\frac{\left(B-2A\right)L^2}{24}\right)\left(1+
\frac{\left(B-4A\right)L^2}{24}+\frac{L^4}{160z_h^4}\right)
\ee
and
\be
\Re S_I=-\frac{\sqrt2L^-}{2\pi{\alpha}^\prime{z_h}^2}\frac{L}{2}
\left(1+\frac{\left(B-2A\right)L^2}{24}\right)\left(1+\frac{BL^2}{24}+
\frac{L^4}{160z_h^4}\right).
\ee
Thus the gauge-gravity prescription (\ref{wilsonconst}) is reduced into
\be
C&=&\left(1-2\Im S_I\right)\left(1-2\left(\Re S_I\right)^2\right) \nonumber\\
&=&\left[1+\frac{L^-L}{\sqrt2\pi{\alpha}^\prime{z_h}^2}\left(1+
\frac{\left(B-2A\right){L}^2}{24}\right)\left(1+
\frac{\left(B-4A\right){L}^2}{24}+\frac{{L}^4}{160z_h^4}\right)
\right] \nonumber\\
&\times& \left[1-\frac{L^{-2}{L}^2}{4{\pi}^2{\alpha^\prime}^2z_h^4}\left(1+
\frac{\left(B-2A\right){L}^2}{12}\right)
\left(1+\frac{B{L}^2}{12}+\frac{{L}^4}{80z_h^4}\right)\right]
\label{IandII}
\ee
Let the first and the second term in the square bracket in the above equation 
(\ref{IandII}) be denoted by I and II, respectively
\be
I& \equiv&\left[1+\frac{L^-L}{\sqrt2\pi{\alpha}^\prime{z_h}^2}\left(1+
\frac{\left(B-4A\right){L}^2}{24}+\frac{{L}^4}{160z_h^4}-
\frac{\left(B-2A\right){L}^2}{24}\right)\right]\nonumber\\
&=&\left[1+\frac{L^-L}{\sqrt2\pi{\alpha}^\prime{z_h}^2}\left(1-
\frac{A{L}^2}{12}+\frac{{L}^4}{160z_h^4}\right)\right] \\
II& \equiv&\left[1-\frac{L^{-2}{L}^2}{4{\pi}^2{\alpha^\prime}^2z_h^4}\left(1+
\frac{\left(B-A\right){L}^2}{6}+\frac{{L}^4}{80z_h^4}\right)\right]
\ee
Therefore the product of the terms I  and II in (\ref{IandII}) yields
\be
C=\left[1-pL-q{L}^2-r{L}^3-s{L}^4-t{L}^5-u{L}^6
+\rm{higher ~~order ~~terms}\right]~,
\ee
where
\be
p& \equiv & - \frac{L^-}{\pi \sqrt{2} \alpha^\prime {z_h}^2} \nonumber\\
q& \equiv &  \frac{{L^-}^2}{4\pi^2 {\alpha^\prime}^2 {z_h}^4} \nonumber\\
r& \equiv &  \frac{AL^-}{12\pi \sqrt{2} \alpha^\prime {z_h}^2} 
\ee
By inverting the equation and ignoring the higher-order
terms, the separation ($L$) is given by
\be
L=\frac{1-C}{p}-\frac{q\left(1-C\right)^2}{p^2}+\frac{\left(1-C\right)^3
\left(2q^2-pr\right)}{p^5}
\ee
Therefore the quenching parameter, $\hat{q}$ is obtained from the 
definition (\ref{qhatourdef}) :
\be
\hat{q} &=&\frac{1}{L^- L^2}\nonumber\\
&=& \frac{L^-}{2 \pi^2 {\alpha^\prime}^2 z_h^4 {(1-C)}^2}
\left[1-\frac{L^-(1-C)}{\sqrt{2} \pi \alpha^\prime z_h^2}-{(1-C)}^2
\left(1+\frac{A\pi^2{\alpha^\prime}^2 z_h^4}{3{L^-}^2}\right) \right],
\ee
which finally results into for $C=\frac{1}{2}$,
\be
\hat{q}= \frac{2L^-}{\pi^2 {\alpha^\prime}^2 z_h^4}\left[\frac{3}{4}-\frac{L^-}{2\sqrt{2}\pi\alpha z_h^2}-\frac{A\pi^2{\alpha^\prime}^2 z_h^4}{12{L^-}^2}\right]
\ee
\noindent {\bf Case II}: In the long-distance limit ($z_m >> z_h$),
let us first express the separation ($L$) as a function of $z_m$. Therefore,
we split up the limits of integration to the equation of motion (\ref{eom}) 
and then integrate it to yield $L$ as a function of $z_m$: 
\be
\frac{iL}{2}&=&\int_0^{z_m}dz\frac{(1+0.5Bz^2)(1-Az^2)}{\sqrt{1-
\frac{z^4}{{z_h}^4}}}\nonumber\\
&=&\int_0^{z_h}dz\frac{(1+0.5Bz^2)(1-Az^2)}{\sqrt{1-\frac{z^4}{{z_h}^4}}}+
\int_{z_h}^{z_m}dz\frac{(1+0.5Bz^2)(1-Az^2)}{\sqrt{1-
\frac{z^4}{{z_h}^4}}}\nonumber\\
&=&1.3z_h+0.15(B-2A){z_h}^3-0.22AB{z_h}^5
+ i\left[1.14z_h+0.5(B-2A)z_m-0.17AB{z_m}^3\right]
\ee
Inverting the series and ignoring the higher-order terms we express 
$z_m$ in terms of $L$ as
\be
z_m=\frac{L-2.28z_h+i2.6z_h}{(B-2A){z_h}^2}
\ee
Now the (renormalized) action (\ref{renoractionlong}) in this limit can be 
expressed as a function of $L$:
\be
S_I&=&-\frac{L^-}{\sqrt{2}\pi \alpha^\prime {z_h}^2}\left[2.44z_h+0.5B{z_h}^2
\left(\frac{L-2.28z_h+i2.6z_h}{(B-2A){z_h}^2}\right)-0.16 i z_h \right.
\nonumber\\
&+& \left. i0.5(B+2A){z_h}^2\left(\frac{L-2.28z_h+i2.6z_h}{(B-2A){z_h}^2}
\right)\right].
\ee
Ignoring the higher-order terms, we get the action as a function of 
$L$, 
\be
S_I \stackrel{z_m \gg z_h}{=}-\frac{L^-}{\sqrt{2}\pi \alpha^\prime {z_h}^2}\left[\frac{0.5BL}{(B-2A)}+i\frac{0.5(B+2A)L}{(B-2A)}\right]
\ee
Now the real and the imaginary parts of renormalized action can be separated,
respectively as
\be
\Re{S_I}&=&-\frac{BL^-L}{2\sqrt{2}\pi (B-2A)\alpha^\prime{z_h}^2} \\
\Im{S_I}&=&-\frac{(B+2A)L^-L}{2\sqrt{2}\pi (B-2A)\alpha^\prime{z_h}^2}
\ee
Thus the gauge-gravity correspondence (\ref{wilsonconst}) in this
limit is translated into:
\begin{eqnarray}
C&=&e^{\left[\frac{(B+2A)L^-L}{\sqrt{2}\pi\alpha^\prime(B-2A)z_h^2}\right]}~
\cos\left[\frac{BL^-L}{\sqrt{2}\pi\alpha^\prime(B-2A)z_h^2}\right]
\label{ccase2}
\end{eqnarray}
Defining 
\be 
a&\equiv &\frac{(B+2A)L^-}{\sqrt{2}\pi\alpha^\prime(B-2A)z_h^2}\nonumber\\
b&\equiv &\frac{BL^-}{\sqrt{2}\pi\alpha^\prime(B-2A)z_h^2}~,
\label{ab}
\ee
the above equation (\ref{ccase2}) has been inverted to give rise the 
expression for the dipole separation ($L$) as
\be
L&=&\frac{C-1}{a}\left[1+\frac{(a^2-b^2)(1-C)}{2a^2}+\frac{(2a^4-3a^2b^2+3b^4){(1-C)}^2}{6a^4} \right. \nonumber\\
&+&\left. \frac{(6a^6-11a^4b^2+16a^2b^4-15b^6){(1-C)}^3}{24a^6}+ \cdot 
\cdot \cdot \cdot \right]
\ee
Using the numerical values of $A$ and $B$ in~\cite{Mia:PRD82'2010} ($A=B=0.124$), the
expressions for $a$ and $b$ in Eq.(\ref{ab}) can be rewritten as 
\be
a=-\frac{3\pi T^2L^-}{\sqrt{2}\alpha^\prime}\quad {\rm{and}} \quad  
b=-\frac{\pi T^2L^-}{\sqrt{2}\alpha^\prime},
\ee
and hence the separation becomes
\be 
L&=&\frac{\sqrt{2}(1-C)\alpha^\prime}{3\pi T^2 L^-}\left[1+\frac{4(1-C)}{9}+
\frac{23{(1-C)}^2}{81}+\frac{301{(1-C)}^3}{1458}+.....\right]
\label{llong}
\ee
Thus the quenching parameter $\hat{q}$ is obtained from (\ref{qhatourdef}) 
by substituting the square of the separation (\ref{llong}) for $C=1/2$
\be
\hat{q}=\frac{102T^4}{{\alpha^\prime}^2} L^-,
\ee
which is seen to be linear in $L^-$.

In the study of DIS on a large nucleus in 
AdS/CFT set up~\cite{Albacete:JHEP0807'2008}, although they did not calculate $\qhat$
directly but if we translate their calculation 
of the saturation scale, $Q_s$ into our calculation we would use 
$\qhat = Q_s^2/L^-$. The way $Q_s$ depends on $L$ is, in turn, dependent 
on which complex branch is chosen. In particular they took $Q_s \sim A^{1/3}
\sim L^-$, since $L \sim A^{1/3}$. Hence in both cases $\qhat$ comes out
 $\sim L^-$, which appears to be in agreement with our calculation.
Since they always assume that $L^- \sim A^{1/3}$ is large so they 
did not keep the inverse powers of $L^-$. Even we checked with their shock-wave
metric~\cite{Albacete:JHEP0807'2008}, where $\qhat$ ($\sim L^-$) for 
large $L^-$.
in terms of the width of shock wave agrees with our result in the respective
limit.

From other perspective of jet quenching phenomena, by comparing the medium 
induced energy loss and the $p_T$-broadening in perturbative QCD with 
that of the trailing string picture of conformal theory
in\cite{Dominguez:NPA811'2008}, they also have used $Q_s \sim L^-$, such 
that $\qhat = Q_s^2/L^- \sim L^-$ is in agreement with everything else we 
obtained so far in our calculations.

\section{Results and Discussions}
We have calculated the quenching parameter, $\hat{q}$ 
in the holographic set-up of gauge-gravity duality, where
the dual gauge theory at finite temperature 
is more closer to thermal QCD than the ${\mathcal{N}}$=4 
SYM theory usually used in the literature.
Moreover we use a more appropriate definition of $\hat{q}$ compatible with 
the strong coupling limit of gauge-gravity duality, for 
which the real part of the Wilson loop expectation value 
is equal to some specific value (1/2). We have found that in both
short and long-distance limit, $\qhat$ depends linearly on $L^-$.
However, in short-distance limit we obtain $1/L^-$ and ${L^-}^2$ correction terms.

It is however worth to mention here that it is not clear what one should do 
with $\hat{q}$ found in a non-perturbative AdS calculation. Since the 
energy loss calculations are usually done using the perturbative 
approximation, one can not simply take a non-perturbative $\hat{q}$
and plug it into the perturbative energy loss expression. But then there 
is nothing else one can do. This is why people calculated drag force on a 
heavy quark without looking for 
$\hat{q}$~\cite{Gubser:PRD74'2006,Gubser:PRD76'2007,Yaffe:JHEP0607'2006} 
or the instantaneous energy loss suffered by light quarks in AdS 
directly~\cite{Ficnar:JPCS446'2013,Martins:JPCS706'2016}.
It would be interesting to see whether the drag calculation would give the 
same $\qhat$ as the one we have obtained. As far as we remember, the drag 
calculation in \cite{Dominguez:NPA811'2008} obtained both $\qhat$ and $Q_s$
which are in qualitative agreement with what we have gotten.

\section{Acknowledgements}
We are grateful to Yuri Kovchegov for his constant and meticulous 
suggestions. It would never have been possible for us to complete
this work without his help. BKP is thankful to the CSIR (Grant No.03 (1215)/12/EMR-II), 
Government of India for the financial assistance.


\begin{thebibliography}{10}
\bibitem{Baier:NPB483'1997}
  R.~Baier, Y.~L.~Dokshitzer, A.~H.~Mueller, S.~Peigne and D.~Schiff,
  Nucl.\ Phys.\ B {\bf 483}, 291 (1997).


\bibitem{Zakharov:JETPLETT65'1997}
  B.~G.~Zakharov,
  JETP Lett.  {\bf 65}, 615 (1997).

\bibitem{Gyulassy:NPB594'2001}
 M.~Gyulassy, P.~Levai and I.~Vitev, Nucl.\ Phys.\ B {\bf 594}, 371 (2001).

\bibitem{Mustafa:APHA22'2005}
M. ~G. ~Mustafa, M. ~H. Thoma, Acta Phys.Hung. A {\bf 22}, 93 (2005).

\bibitem{Bruke:PRC90'2014} K.~M.~Burke et al., Phys. Rev. C {\bf 90},
014909. (2014).
Karen M. Burke,1 


\bibitem{Maldacena:PRL80'1998}
 J.~M.~Maldacena, Phys.\ Rev.\ Lett.\  {\bf 80}, 4859 (1998).

\bibitem{Gubser:PLB428'1998}
 S.~S.~Gubser, I.~R.~Klebanov and A.~M.~Polyakov, 
Phys.\ Lett.\ B {\bf 428}, 105 (1998).

\bibitem{Witten:ATMP2'1998} E.~Witten,
 Adv.\ Theor.\ Math.\ Phys. {\bf 2}, 253 (1998);
{\bf 2}, 505 (1998).

\bibitem{Son:ARNPS57'2007}
 D.~T.~Son and A.~O.~Starinets,
 Ann.\ Rev.\ Nucl.\ Part.\ Sci. {\bf 57}, 95 (2007). 

\bibitem{Liu:PRL97'2006}
 H.~Liu, K.~Rajagopal, and U.~A.~Wiedemann, 
Phys.\ Rev.\ Lett. {\bf 97}, 182301 (2006).

\bibitem{AHMueller:NPB335'1990} A. ~H. ~Mueller, Nucl.Phys. B{\bf 335},115 
(1990).

\bibitem{Casalderrey:NPA'2014} J.~Casalderrey-Solana, D. ~C. ~Gulhan, J. ~G.~ Milhano, D.~
Pablos, K.~Rajagopal, Nucl. Phys. A{\bf 932}, 421 (2014); 
JHEP {\bf 1410}, 19 (2014), Erratum: JHEP {\bf 1509}, 175 (2015). 

\bibitem{Eramo:PRD84'2011} F.~D'Eramo, H.~Liu, K.~Rajagopal,
Phys. Rev. D {\bf 84} (2011) 065015.

\bibitem{Abir:PLB748'2015}
 R.~Abir, Phys.\ Lett.\ B {\bf 748}, 467 (2015).

\bibitem{Albacete:JHEP0807'2008} 
J.~L.~Albacete, Y.~V.~Kovchegov, A.~Taliotis, JHEP {\bf 0807}, 074 (2008). 

\bibitem{Kovchegov:Cambridge'2012} 
Y.~V.~Kovchegov, E.~Levin, 
Quantum Chromodynamics at High Energy, (Cambridge Monographs on Particle 
Physics, Nuclear Physics and Cosmology), 1st Edition (2012).
 
\bibitem{Klebanov:JHEP0008'2000}
 I.~R.~Klebanov and M.~J.~Strassler, JHEP {\bf 0008}, 052 (2000).

\bibitem{Ouyang:NPB699'2004} P.~Ouyang, Nucl.\ Phys.\  B {\bf 699}, 207 (2004).


\bibitem{Maldacena:PRL86'2001}
J.~M.~Maldacena and C.~Nunez, Phys.\ Rev.\ Lett.\  {\bf 86}, 588 (2001).

\bibitem{Mia:PRD82'2010} M.~Mia, K.~Dasgupta, C.~Gale and S.~Jeon,
Phys.\ Rev.\ D {\bf 82}, 026004 (2010).

\bibitem{Binoy:PRD91'2015}
 Binoy Krishna Patra and H.~Khanchandani,
 Phys.\ Rev.\ D {\bf 91}, 066008 (2015).

\bibitem{Binoy:PRD92'2015}
 Binoy Krishna Patra, H.~Khanchandani, and Lata Thakur, 
Phys.\ Rev.\  D {\bf 92}, 085034 (2015).


\bibitem{Klebanov:NPB578'2000}
 I.~R.~Klebanov and A.~A.~Tseytlin,
 Nucl.\ Phys.\  B {\bf 578}, 123 (2000).

\bibitem{Mia:NPB839'2010} 
M.~Mia, K.~Dasgupta, C.~Gale and S.~Jeon,
Nucl.\ Phys.\ B {\bf 839}, 187 (2010).

\bibitem{Mia:PLB694'2011}
 M.~Mia, K.~Dasgupta, C.~Gale and S.~Jeon,
Phys.\ Lett.\ B {\bf 694}, 460 (2011).

\bibitem{Chen:PRD87'2013}
 F.~Chen, L.~Chen, K.~Dasgupta, M.~Mia, and O.~Trottier,
  Phys.\ Rev.\ D {\bf 87}, 041901 (2013).

\bibitem{Sakai:PTP113'2005}
T.~Sakai and S.~Sugimoto,
 Prog.\ Theor.\ Phys. {\bf 113}, 843 (2005).

 \bibitem{Dasgupta:JHEP1507'2015}
  K.~Dasgupta, C.~Gale, M.~Mia, M.~Richard, O.~Trottier, JHEP {\bf 1507}, 122 (2015).


\bibitem{Karch:JHEP0206'2002} A.~ Karch, E.~Katz, JHEP {\bf 0206}, 043 (2002). 

\bibitem{Sakai:JHEP0309'2003} T.~Sakai, J.~Sonnenschein, 
JHEP {\bf 0309}, 047 (2003).

\bibitem{Dominguez:NPA811'2008}
 F.~Dominguez, C. ~Marquet, A.~H.~ Mueller, B.~ Wu, B.~W.~ Xiao
, Nucl.\ Phys.\ A {\bf 811}, 197 (2008).


\bibitem{Armesto:JHEP0609'2006} N.~Armesto, J.~D.Edelstein, J.~Mas,
JHEP {\bf 0609},039 (2006).

\bibitem{Ficnar:JPCS446'2013} A.~Ficnar, J.~Noronha, M.~Gyulassy
J.\ Phys.\ Conf.\ Ser. {\bf 446}, 012002 (2013).

\bibitem{Martins:JPCS706'2016} S.~Martins, C.~B.Mariotto,
J.\ Phys.\ Conf.\ Ser. {\bf 706}, 052035 (2016).


\bibitem{Arnold:JHEP1302'2013} P.~Arnold, P.~Szepietowski, D.~Vaman, G.~Vong,
JHEP {\bf 1302}, 130 (2013).

\bibitem{Arnold:JHEP1207'2012} P.~Arnold, P.~Szepietowski, D.~Vaman, JHEP 
{\bf 1207}, 024 (2012).


033 (2011).  

\bibitem{Gubser:PRD74'2006} S. S. Gubser, Phy. Rev. D {\bf 74}, 126005 (2006).

\bibitem{Gubser:PRD76'2007} S. S. Gubser, Phy. Rev. D {\bf 76}, 126003 (2007).

\bibitem{Yaffe:JHEP0607'2006} C.P. Herzog, A. Karch, P. Kovtun, C. Kozcaz,
and L.G. Yaffe, JHEP {\bf 0607}, 013 (2006).

\bibitem{Li:arXiv:1605.00188} S.~Li, K.~A.Mamo, H.~Yee, arXiv:1605.00188[hep-ph].


\bibitem{Lin:PLB641'2006} Feng-Li Lin, Toshihiro Matsuo, Phys.Lett. B641 (2006) 45-49 

\end{thebibliography}
\end{document}